\documentclass{phbauth}
\usepackage{graphicx}

\begin{document}

\begin{frontmatter}

\title{Finite Range Model Interaction Potential for d-wave
Superconductors: $T_{\rm c}$ vs.\ Doping in the Cuprates}

\author[address1]{Jorge Quintanilla\thanksref{thank1}},
\author[address1]{Balazs L.\ Gy\"orffy}

\address[address1]{H.\ H.\ Wills Physics Laboratory, University of Bristol, Tyndall Avenue, Bristol BS8 1TL, U.K.}

\thanks[thank1]{Corresponding author. Fax: +44 (0)117 925 5624. \mbox{E-mail}: J.Quintanilla@bristol.ac.uk}

\date{12 July 1999}

\begin{abstract}

We study a simple, BCS like, model which describes unconventional superconductivity on the basis of an electron-electron attraction corresponding to the delta-shell potential: $U(r_{12}) = -g \delta(r_{12}-r_0)$. It predicts a $T_{\rm c}$ vs.\ doping behavior similar to that characteristic of the High $T_{\rm c}$ cuprates.

\end{abstract}

\end{frontmatter}

The simplest, useful model of superconductivity features an attractive pair potential $U^{\rm BCS}(r_{12})=-u \delta({\bf r}_{12})$ and an energy cutoff $\hbar \omega_D$ \cite{BCS,Gor'kov}. Although the coupling constant $u$ and the frequency $\omega_D$ can be chosen to yield a transition temperature $T_{\rm c} \sim 100 {\rm K}$ it can not be used to discuss the high $T_{\rm c}$ cuprates because it, inevitably, leads to s-, and not d-wave pairing \cite{Annett}. In this note we propose a slight modification of the above model and demonstrate that it can describe Cooper pairs with any desired internal angular momentum, while remaining almost as tractable as the original. Surprisingly, the new model implies a rise and fall of $T_{\rm c}$ with the carrier density $n$ and can fit the well-known, apparently universal, empirical relation between $T_{\rm c}$ and doping $n$, characteristic of the high $T_{\rm c}$ cuprates \cite{Tallon}.

In a BCS-like theory, for free electrons interacting with a general attractive pair potential $U(r_{12})$, the Hartree-Fock-Gor'kov approximation yields a gap equation for $\Delta({\bf k})=\sum_{\rm lm}\Delta_{\rm lm}(k)Y_{\rm lm}(\hat{k})$, where $Y_{\rm lm}$ is a spherical harmonic. At $T_{\rm c}$, it is
      \begin{eqnarray}
      \lefteqn{\Delta_{l,m}(k) =} \\
      &=&-\int_{0}^{\infty} dk' k'^2
      \frac{K_l(k',k)}{2\epsilon({\bf k'})}
      \tanh \left[ \frac{\epsilon({\bf k'})}{2k_BT_c} \right]
      \Delta_{l,m}(k'), \nonumber
      \end{eqnarray} 
where $\epsilon(k)$ is the normal-state dispersion relation, measured from the Fermi energy, and the kernels are
	\begin{equation}
	K_{\rm l}(k',k) = \frac{2}{\pi}
	\int_0^{\infty} {\rm d}r \
	j_{\rm l}(k'r) \ r^2 U(r) \ j_{\rm l}(kr).
	\label{kernel}
	\end{equation}
Evidently, any interaction that reduces to a narrow peak centred at $r_{\rm 12}=0$ favours s-wave superconductivity (because $j_{\rm l}(0) = 0$ for all $l>0$). This is the case of the BCS interaction
      \(
      U^{\rm BCS}(r_{12})=-u \delta({\bf r}_{12}),
      \)
giving $\Delta = \mbox{constant}$.
Thus, to account for d-wave pairing, it is natural to assume
 that, in the cuprates, the peak of $U(r_{12})$ is at some finite distance $r_0$ from the origin.
 
 The simplest such interaction is the delta-shell potential \cite{Gottfried,Villarroel},
      \begin{equation}
      U(r_{12})=-g \delta(r_{12}-r_0),
      \label{delta-shell}
      \end{equation}
for which
	\(
	\Delta_{\rm lm}(k) \propto j_{\rm l}(k r_0).
      \label{radialpart}
	\)
Unlike in the case of $U^{\rm BCS}(r_{\rm 12})$, for this potential the gap equation converges without the need to introduce a cutoff. 

\begin{figure}[btp]
\begin{center}\leavevmode
\includegraphics[width=0.9\linewidth]{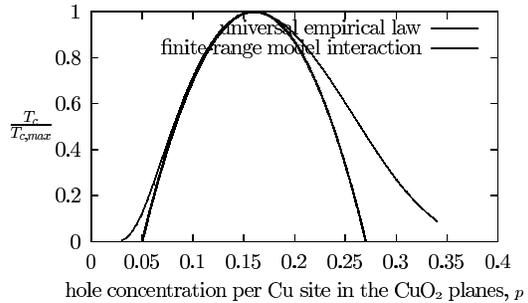}
\caption{The best fit of the $T_{\rm c}/T_{\rm c,max}$ vs.\ hole doping curve predicted by the delta-shell model to the empirical universal law.}\label{figure}\end{center}\end{figure}

A surprising, and interesting feature of this modified BCS model is that the interplay between the new microscopic length scale, $r_0$, and the Fermi wavelength, $k_{\rm F}^{-1}$, leads to a rise and fall of $T_{\rm c}$ as a function of carrier concentration, $n$, as $k_{\rm F}$ approaches a maximum or a node of the gap function, $\Delta_{\rm lm}(k)$, respectively.
Experimentally, the critical temperature of the cuprates obeys, approximately, the following law \cite{Tallon}:
      \begin{equation}
      T_c/T_{c,max} = \left[ 1-82.6(n-0.16)^2 \right]
      \label{empirical}
      \end{equation}
where $n$ is the number of holes per copper site in the CuO$_2$ planes. For d-wave solutions to the delta-shell model, the values of the two parameters in (\ref{delta-shell}) that produce the best fit to this law (Fig.\ \ref{figure}) are
      \begin{eqnarray}
      r_0 &=& 2.2 \times r_{\rm Cu-Cu} \label{r0} \\
      g/r_0 &=& 0.6 \times \hbar^2/2m^{*}r_{\rm Cu-Cu}^2 \label{loc}
      \end{eqnarray}
Here $r_{\rm Cu-Cu}$ is the \emph{average} distance between superconducting copper sites, and $\hbar^2/2m^{*}r_{\rm Cu-Cu}^2$ is the corresponding localisation energy ($m^{*}$ is the effective mass of a hole). Note that this result implies a relationship between $g$ and $r_0$ that is the same for all materials from which the phenomenological law (\ref{empirical}) has been deduced.

We can now calculate the \emph{absolute} value of $T_{c}$. This is
      \(
      T_{\rm c,max} =
      0.08 \times \hbar^2/2m^{*}r_{\rm Cu-Cu}^2,
      \)
which is of the right size. For example, for YBCO we get $T_{\rm c} = 186K (m_{\rm e}/m^{*})$, which equals $92K$ if we take $m^{*}=2.0 \ m_{\rm e}$.

To summarise, we can account for d-wave symmetry, high-$T_{\rm c}$ and the universal $T_{\rm c}$ vs.\ doping law in the cuprates if we assume that the attractive effective interaction between holes acts at some finite distance, $r_0$. Reassuringly, from the point of view of the physical content of the model, $r_0$ works out to be close to the average separation between the copper sites of the CuO$_2$ planes. Furthermore, the strength of the interaction, $g$, equals $2.9 \times \hbar^2/2m^{*}r_0^2$ for all materials. Clearly, finding the physical mechanism that gives rise to an attractive interaction with these features remains an outstanding challenge.

\begin{ack}
We wish to thank Jonathan Wallington and James Annett for many useful discussions.

 This work has been supported by EU TMR grant number \mbox{ERBFMBICT983194}. 

  BLG would like to thank the Erwin Schr\"odinger Institute for Mathematical Physcis for hospitality during the preparation of the manuscript.
\end{ack}

\end{document}